\documentclass[lettersize,journal]{IEEEtran}
\usepackage{amsmath, amsfonts, amssymb}
\usepackage{array}
\usepackage[caption=false,font=normalsize,labelfont=sf,textfont=sf]{subfig}
\usepackage{textcomp}
\usepackage{stfloats}
\usepackage{url}
\usepackage{verbatim}
\usepackage{graphicx}
\usepackage{physics}
\usepackage{xcolor}
\usepackage{multicol}
\usepackage[ruled,lined,linesnumbered]{algorithm2e}
\usepackage[style=ieee]{biblatex}
\usepackage{nccmath}
\usepackage{soul}
\usepackage{flushend}

\usepackage{soul}

\newcommand*\R{\mathbb{R}}
\newcommand*\C{\mathbb{C}}

\begin{document}
\title{Grouping of $N-1$ Contingencies for Controller Synthesis: A Study for Power Line Failures}
\author{Neelay Junnarkar, Emily Jensen, Xiaofan Wu, Suat Gumussoy, Murat Arcak
\thanks{This work was supported in part by Siemens Corporation R{\&}D funding and by NSF grant CNS-2135791.}
\thanks{Neelay Junnarkar, Emily Jensen, and Murat Arcak are with the department of Electrical Engineering and Computer Sciences at the University of California, Berkeley. (emails: $\{$neelay.junnarkar, emilyjensen, arcak$\}$ @berkeley.edu)}
\thanks{Xiaofan Wu and Suat Gumussoy are with Siemens Technology. (emails: $\{$xiaofan.wu, suat.gumussoy$\}$ @siemens.com}
}


\maketitle

\begin{abstract}

The problem of maintaining power system stability and performance after the failure of any single line in a power system (an ``$N-1$ contingency'') is investigated. Due to the large number of possible $N-1$ contingencies for a power network, it is impractical to optimize controller parameters for each possible contingency a priori. A method to partition a set of contingencies into groups of contingencies that are similar to each other from a control perspective is presented. Design of a single controller for each group, rather than for each contingency, provides a computationally tractable method for maintaining stability and performance after element failures. The choice of number of groups tunes a trade-off between computation time and controller performance for a given set of contingencies. Results are simulated on the IEEE 39-bus and 68-bus systems, illustrating that, with controllers designed for a relatively small number of groups, power system stability may be significantly improved after an $N-1$ contingency compared to continued use of the nominal controller. Furthermore, performance is comparable to that of controllers designed for each contingency individually. 


\end{abstract}


\section{Introduction}
\IEEEPARstart{T}{he}
robustness of a power system against disturbances is important for power system stability. To this end, controller parameters are tuned to minimize  inter-area oscillations due to disturbances. Some of the available approaches are pole placement \cite{othman1989}, $\mathcal{H}_2$ control \cite{wu2016,zolotas2007, li_grid_2019, dehghani_control_2022}, and $\mathcal{H}_\infty$ control \cite{SCHULER2014379, MESANOVIC2020104490, dehghani_control_2022}.
These parameters are tuned for a particular power system and may be sub-optimal if the dynamics of the power system change, due to, e.g., changes in operating conditions or line failures. Existing approaches to this problem include robust control design methods which enable performance and stability for a single controller in feedback with any plant in a fixed set of plants  \cite{MESANOVIC2020104490, xue2022}.

In this work, we consider the problem of maintaining stability and performance of a power network when dynamics change due 
to so-called $N-1$ contingency events, as stability in the presence of these events is required by the  North American Electricity Reliability Corporation (NERC) \cite{nerc}.
 An $N-1$ contingency event is the failure of a single element in a power system, such as
 generators, transmission lines, and transformers. In what follows, we restrict our attention to line failures.
 Line failures change the topology of a power system, and thus the dynamics. A controller that optimally rejects disturbances for the nominal system may 
perform suboptimally on the perturbed system.

Related to this problem is the identification of problematic contingencies or critical components
\cite{huang2021}. Existing results include methods to identify contingencies that initiate cascading failure \cite{eppstein2012}, to identify contingencies that involve several failing components \cite{chen2005}, and to estimate the probabilities of various types of contingencies \cite{jiang_contingency_2021}. Identification of critical $N-1$ contingencies has also been considered in parallel with controller design \cite{xue2022}.


In this work, we 
focus on computational tractability in the setting of
maintaining stability and performance of a power network in the presence of changes due to 
$N-1$ contingencies.
To optimize performance, a separate controller should be synthesized for each  possible contingency with each new operating condition. 
This should be done in advance of failure as designing a controller after a contingency has already occurred leaves the power system vulnerable to disturbances and oscillations during the controller design period. 
Although this is optimal for performance, it is computationally intractable
due to the number of contingencies, and the time-scale at which operating conditions (and thus dynamics) of power networks change. 
{
On the opposite extreme, a single controller could be synthesized for use in the event of \emph{any} failure scenario. This is impractical as it will likely lead to instability or at least poor performance in a variety of failure scenarios.
Intermediate to these extremes, we propose partitioning the space of failure scenarios into $n$ ``groups", and synthesizing a controller for each such group.
The choice of $n$ serves to tune the tradeoff between performance and computational tractability: $n$ should be larger than one and much smaller than the total number of failure scenarios.
Designing one controller per group reduces the number of controllers that need to be synthesized to handle all contingencies since there need only be as many controllers as there are groups. When a contingency occurs, the appropriate group controller can be applied immediately since all controllers are designed in advance.
}

{
While existing methods in robust control can be applied to synthesize a controller 
to stabilize the multiple systems within each group, a partitioning of the space of contingencies into the ``best" groups for controller synthesis is unknown. The aim of this work is to provide a method to find a well-performing grouping, in a computationally efficient manner.
}
In power systems literature, clustering techniques have been commonly used in areas related to decomposing power systems into voltage control areas \cite{satsangi2011, wu2018, verma2016, jayawardene2020}. The clustering use most closely related to our work is for identifying a small subset of contingencies to use during design as representative of the types of power system dynamics that may occur \cite{sigrist2008}.

{
Relevant to our work is \cite{xue2022}, where the authors present a method to identify a critical subset of $N-1$ contingencies, and a method to optimize a controller to be robust to the occurrence of these select contingencies.
}
{
Our work is advantageous compared to some aspects of \cite{xue2022}, and complementary to others.
First, whereas \cite{xue2022} identifies and handles only a subset of contingencies, our proposed method of partitioning contingencies handles all contingencies.
Second, whereas \cite{xue2022} designs a single controller to handle all (in general, dissimilar) critical contingencies, our proposed method designs one controller per group, where each group consists of contingencies resulting in similar dynamics.
Designing a single controller for a set of dissimilar contingencies as in \cite{xue2022} limits the performance that the controller can achieve on any one contingency. If the dynamics of the contingencies are sufficiently different, there may not exist a single controller which stabilizes all selected contingencies.
Our proposed method avoids this problem by constructing the groups such that all contingencies in a group result in similar dynamics, and designing one controller per group.
We note that the control design technique presented in \cite{xue2022} could be used as a component of our approach to design controllers for each group.
}

{
The main results of this work are as follows:}
\begin{itemize}
{
 \item  A method for partitioning a set of contingencies for a power network into groups that are similar to each other from a control perspective. This proposed grouping method additionally serves as an analysis tool to identify severe contingencies.
}
{
    \item A procedure that leverages this grouping method to design controllers to handle all contingencies in a computationally tractable manner.
    \item Experimental verification of the effectiveness of this grouping and control procedure through simulation using $\mathcal{H}_\infty$ and $\mathcal{H}_2$ performance measures.
}
 \end{itemize}

The rest of this paper is organized as follows. In Section~\ref{sec:modeling} we describe the power system model and performance metric we consider. In Section~\ref{sec:grouping} we present our main contributions: a method for grouping contingencies based on a distance metric together with a clustering algorithm{, and a procedure to use this grouping method to design controllers to handle all contingencies}. In Section~\ref{sec:simulation} we
illustrate the performance of our method through simulation on the IEEE 39-bus and 68-bus systems.
In Section~\ref{sec:conclusion} we present our conclusions and directions for future work.

{
\subsection{Notation}
The $\mathcal{H}_\infty$ and $\mathcal{H}_2$ norms of a transfer function $T$ are denoted by  $\|T\|_{\mathcal{H}_\infty}$ and $\|T\|_{\mathcal{H}_2}$, respectively. These norms are defined in Section~\ref{subsec:perf-obj}.
The transfer function from disturbance to performance output of the closed-loop system of a plant $P$ and controller $K$ is denoted by $\mathcal{F}(P, K)$.
}

\section{Problem Set-up} \label{sec:modeling}

\subsection{Power System Model}

A power system can be modeled by a system of nonlinear differential-algebraic equations (DAEs). The differential equations describe the dynamics of, for example, the generators, while the algebraic equations represent the network equations. We consider a linear time-invariant (LTI) model where the DAE has been linearized around an operating point and the network equations have been eliminated.
We model the operating point as fixed for some period of interest, noting that it will change on a slow time scale as conditions such as loads throughout the network change.
{
Note that this LTI model can be formed from the DAE model regardless of whether the power system has synchronous generators, inverter-based resources (such as renewable energy sources), or a mix of the two. See, for example, \cite{sajadiPlaneWaveDynamic2024} and \cite{baracModelingInitializationVirtual2021}.
}

{
We assume the LTI model is of the form
\begin{equation} \label{eq:open-loop-dynamics}
\dot{x} = Ax + B_w w + B_u u,
\end{equation}
where \(x \in \R^n\) is the deviation of the power system state from the operating point, $w \in \R^{n_w}$ is the disturbance, and \(u \in \R^m\) is the control input.}

{
In a power system with synchronous generators, for example, the LTI model can be constructed such that $x$ consists of the following: $\theta$, the vector of generator rotor angles, with the average of rotor angles removed; $\dot{\theta}$, the vector of deviations of generator rotor frequencies from the operating frequency; and additional states accounting for fast electrical dynamics \cite{wu2016}.
}
{
An LTI model with this interpretation can also be constructed for power systems with inverter-based resources using virtual inertia techniques. A review of such techniques can be found in \cite{yap_virtual_2019}.
}



The contingencies we consider are single-line failures. For a power system, we can construct a graph where buses are nodes and lines between buses are edges. A line failure is represented by the removal of the corresponding edge from the graph.
For simplicity of modeling, we consider line failures that do not disconnect the graph.
This ensures that the states of the model have the same interpretation across all contingencies.
Under these assumptions, only the \(A\) matrix differs between contingencies.
Thus, the dynamics of contingency $i$ can be represented as
\begin{equation} \label{eq:open-loop-dynamics-i}
\dot{x} = A_i x + B_w w + B_u u.
\end{equation}
{
Note that under contingency $i$, the state vector $x$ represents the deviation of the power system state from the operating point of contingency $i$.

With this model, the event where contingency $i$ occurs is where the dynamics change from nominal dynamics $(A_0, B_w, B_u)$, to contingency dynamics $(A_i, B_w, B_u)$, and the operating point changes from the nominal operating point to the operating point under contingency $i$.
Since we handle only single-line failures, we assume that this change in dynamics happens once.
}

{
\subsection{Controller Model}
We model the controller as a state-feedback LTI controller.
In particular, for a a plant $P$ as in equation~\eqref{eq:open-loop-dynamics}, the controller $K$ is of the form 
\begin{equation} \label{eq:controller}
    \begin{aligned}
        \dot{x}_k & = A_k x_k + B_{k x} x \\
        u & = C_k x_k + D_{k x} x
    \end{aligned}
\end{equation}
where $x_k \in \mathbb{R}^{n_k}$ is the controller state, $x$ is the state of the plant $P$, and $u$ is the control input to plant $P$. Matrices $A_k$, $B_{k x}$, $C_k$, and $D_{k x}$ are all tunable controller parameters.

This is a centralized controller, with all entries of the control input $u$ being a function of all entries of the plant state $x$. Further, we assume this centralized communication topology does not change when a line failure occurs; i.e. line failure does not result in communication failure.

With this controller, the closed-loop system of $P$ and $K$ can be written as
\begin{equation} \label{eq:closed-loop}
    \begin{aligned}
        \dot{\tilde{x}} &= \mathcal{A} \tilde{x} + \mathcal{B}_{w} w
    \end{aligned}
\end{equation}
where
\begin{align*}
    \mathcal{A} = \begin{bmatrix}
        A + B_u D_{k x} & B_u C_k \\
        B_{k x} & A_k
    \end{bmatrix},
    &&
    \mathcal{B}_w = \begin{bmatrix}
        B_w \\ 0
    \end{bmatrix}.
\end{align*}

The control design problem is to find suitable parameters $A_k$, $B_{k x}$, $C_k$, and $D_{k x}$, where we define ``suitable'' with respect to a performance objective, defined in the next subsection.
}

\subsection{Performance Objective} \label{subsec:perf-obj}

{
We consider the control problem of suppressing effects of disturbances.
In our experiments in particular, we consider the effect of disturbances on inter-area oscillations.
To measure performance of a controller, we first add a virtual output $z$ to the closed-loop system, as follows, for some $\mathcal{C}$ and $\mathcal{D}$.
\begin{equation} \label{eq:closed-loop-perf}
    \begin{aligned}
        \dot{\tilde{x}} &= \mathcal{A} \tilde{x} + \mathcal{B}_{w} w \\
        z &= \mathcal{C} \tilde{x} + \mathcal{D} w
    \end{aligned}
\end{equation}
We refer to the transfer function from disturbance $w$ to performance output $z$ of this closed-loop system of plant $P$ and controller $K$ by $\mathcal{F}(P, K)$.
Then, we measure disturbance suppression using the 
$\mathcal{H}_\infty$ norm of $\mathcal{F}(P, K)$, where a smaller norm means more suppression, with the optimal value being $0$. The $\mathcal{H}_\infty$ norm is defined as 
\begin{equation} \label{eq:hinf-norm}
    \|G\|_{\mathcal{H}_\infty} \triangleq \mathrm{ess}\sup_{\omega \in \R} \bar{\sigma}[G(j\omega)] \\
\end{equation}
where $\bar{\sigma}$ denotes the maximum singular value of a matrix \cite{zhou_robust_1996}.
This metric represents the maximum possible amplification of an input disturbance. Methods involving  $\mathcal{H}_\infty$ control are widespread in power systems control literature, e.g. in \cite{xue2022, raoufat_virtual_2016, zhu_robust_2003}.

}
{
In this work, we focus on control design and performance measurement using the $\mathcal{H}_\infty$ norm. However, we note that this norm can be replaced by other norms, as long as the same norm is used consistently throughout. Thus, in our experiments, we will also consider the $\mathcal{H}_2$ norm. The $\mathcal{H}_2$ norm is defined as
\begin{equation}\label{eq:h2-norm}
    \|G\|_{\mathcal{H}_2} \triangleq \sqrt{\frac{1}{2\pi}\int_{-\infty}^\infty \mathrm{Trace}[G^\star(j\omega)G(j\omega))] d\omega}.
\end{equation}
In the case where the disturbances $w$ are stochastic, the squared  $\mathcal{H}_2$ norm corresponds to the variance in performance output $z$ at steady-state \cite{wu2016}.
}

{
For our experiments, we use a performance output previously used in \cite{wu2016} for suppressing inter-area oscillations. 
This performance output is defined with respect to inertia, which may be either inertia of a synchronous generator, or virtual inertia of a virtual synchronous generator.
}
In particular, let \(\tilde{\theta}\) be the deviation of the generator rotor angles from the average angle, \(\dot{\theta}\) be the deviation of rotor frequencies from the operating frequency, and \(M_G\) be the diagonal matrix of generator inertias. Then we define the performance output of the system to be
\[
z = \mqty[I & 0 \\ 0 & (\frac{1}{2}M_G)^{1/2}] \mqty[\tilde{\theta} \\ \dot{\theta}] .
\]




\section{Grouping} \label{sec:grouping}

In the event of a contingency, one might continue to use a controller designed for the nominal (pre-contingency) system. However, there is no guarantee that this nominal controller will perform well on the perturbed system. To achieve optimal performance, one controller would be designed for each possible contingency and then be applied in the event that that contingency occurs. However, this is computationally intractable for large numbers of contingencies, especially since each controller would need to be updated with each significant change in operating conditions of the network. Therefore, we present a grouping method to partition a set of $M$ contingencies into $k$ groups and to design one controller per group of contingencies.
{For a group of contingencies, the same controller is applied no matter which contingency occurs.} 
If a contingency occurs, its group is determined, and then the group's controller is applied, replacing the nominal controller.

\begin{figure}[t]
    \centering
    \includegraphics[width=0.9\linewidth]{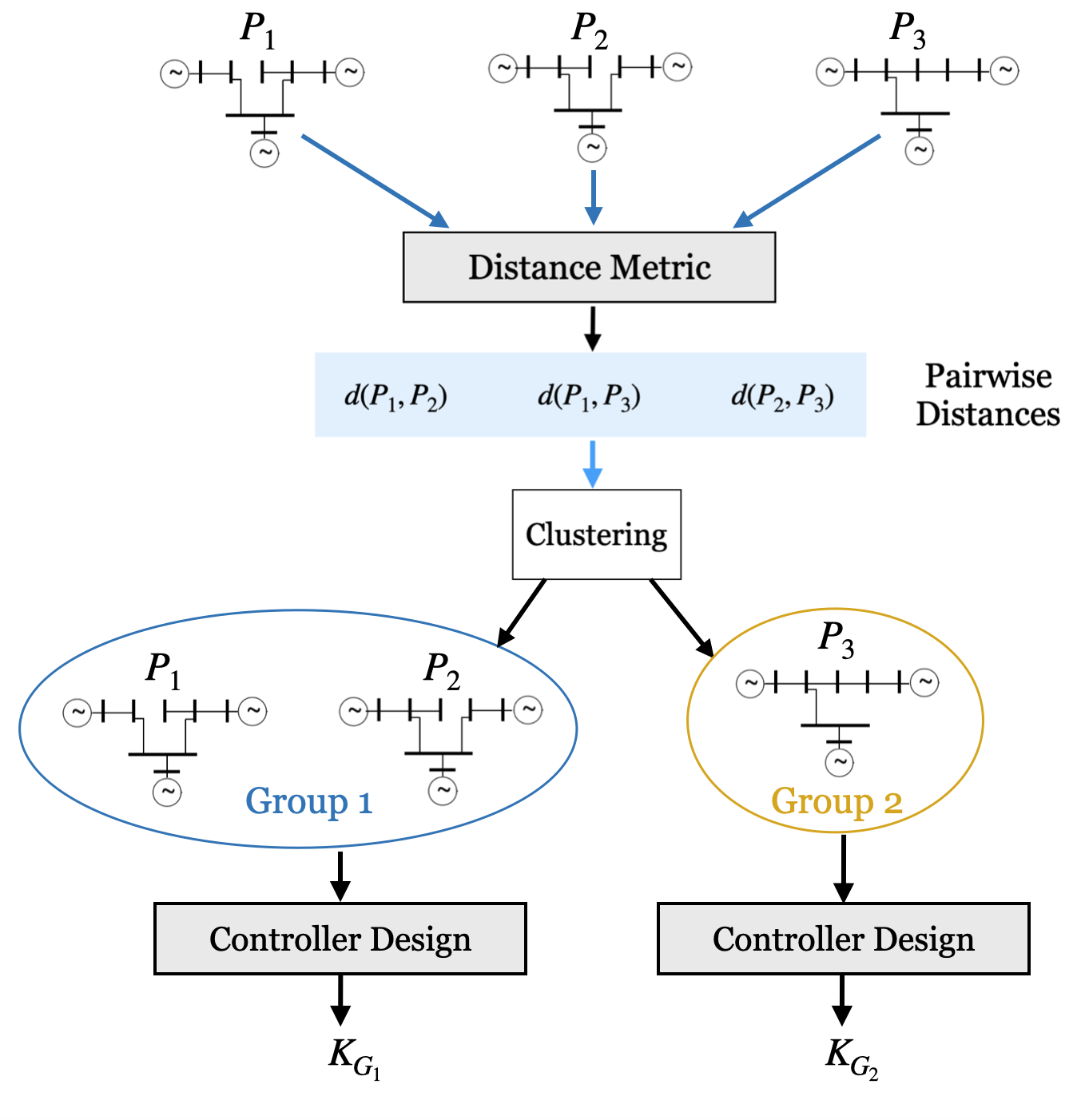}
    \caption{Illustration of this grouping method on a 3 generator system considering 3 possible line failures, and grouping into 2 groups. Pairwise distances between perturbed systems are computed, followed by clustering of the perturbed systems, and finally controller design for each group of systems. }
    \label{fig:grouping-diagram}
\end{figure}

This grouping procedure is depicted in Figure~\ref{fig:grouping-diagram} for a 3-generator system. For illustration, we consider three $N-1$ contingencies of interest, resulting in the three perturbed systems denoted by $P_1,~P_2$, and $P_3$. These three contingency scenarios are grouped by (i) computing a pairwise distance between all perturbed systems of interest and (ii) clustering these systems according to these pairwise distances. A controller design method is applied to obtain one single controller to use in feedback with all systems in each given group. In what follows, we describe various pairwise distance metrics and clustering algorithms.

Many controller design approaches can be utilized once the groups have been formed. 
Here, we consider 
finding the single controller $K$ that minimizes $\max_i \|\mathcal{F}(P_i, K)\|_{\mathcal{H}_\infty}$ for a group of plants $P_1, \dots, P_M$. This minimizes the worst-case $\mathcal{H}_\infty$ norm that can occur when applying controller $K$ to any plant in the group and is used as a proxy for minimizing the worst-case oscillations resulting from any system in the group.
{Note that the control design approach applied to each group can be freely modified to handle other power system considerations, such as small signal stability and transient response.
}

We formulate the grouping problem as a clustering problem with two modules: a metric for measuring distance between contingencies, and a clustering algorithm for computing a partition. The metric should be chosen with respect to the controller design method such that if two contingencies are close to each other, then a controller that achieves high performance on one of the contingencies should achieve high performance on the other. The clustering algorithm uses the pairwise distances to group contingencies. Factors affecting clustering algorithm choice may include the specific controller design technique being used and desired performance criteria. For example, some robust control design techniques on a group of plants depend on the maximum distance between any pair in the group, so a clustering algorithm that minimizes this distance would be desirable. Finally, for any combination of metric and clustering algorithm, the number of groups can be tuned to trade-off performance and computation time.

Algorithm~\ref{alg:alg} summarizes the process of creating and using a library of controllers to handle a set of contingencies.
{
To account for changing load and generation levels over time, the ``offline'' grouping and controller design process can be run repeatedly, with the next iteration beginning as soon as the previous iteration ends.
The ``online'' portion of the algorithm, run in parallel, then uses the most up-to-date set of contingencies, groups, and controllers. 
}

\begin{algorithm}[b]
\SetKw{Function}{Function}
\SetKwBlock{FnBody}{is}{end}
\caption{Grouping and Controller Selection}
\label{alg:alg}
\KwIn{Open-loop dynamics of each contingency $P_1, \dots, P_M$, number of groups $k$.}

\tcp{Offline}
$D \gets$ form matrix of pairwise distances from $\{P_i\}_{i=1}^M$\;
$G_1, \dots, G_k \gets$ cluster $\{P_i\}_{i=1}^M$ into $k$ groups using $D$\;
\For{$i \in \{1, \dots, k\}$}{
    $K_i \gets$ design controller for $G_i$\;
}

\tcp{Online}
$K \gets$ nominal controller\;

\If{contingency $P$ occurs}{
    find $i$ such that $P \in G_i$\;
    $K \gets K_i$\;
}

\end{algorithm}

\subsection{Metrics}

We investigate three metrics to measure distance between contingencies.  Due to the number of pairwise distances scaling in the square of the number of contingencies, we choose metrics with reasonable computation times. We apply these metrics to the plants in feedback with the nominal controller. A metric may be refined for the needs of a particular application. For example, if an application is sensitive to low-frequency differences between plants, one could apply a low-pass filter to the frequency response creating a weighted frequency response metric.


\subsubsection{Frequency Response (FR)}

Let the transfer functions from disturbance input $w$ to performance output $z$ of two LTI systems be $T_{wz,1}(j\omega)$ and $T_{wz,2}(j\omega)$. For a given set of frequencies \(\omega_1, \dots, \omega_n\), we sample both $T_{wz,1}$ and $T_{wz,2}$ at these same points and vectorize the result to form the vectors $f_1, f_2 \in \C^{n \cdot n_w \cdot n_z}$. Then, we define the frequency response metric as 
\[
d_f(f_1, f_2) = \|f_1 - f_2\|_2
\]
using the usual \(\ell^2\) norm.

Note that similar metrics could be defined using $L^p$ norms directly on $T_{wz,1}$ and $T_{wz,2}$, but these are significantly slower to compute than a sampling-based approach.

\subsubsection{Step Response (SR)}

We define a step response based metric largely following our definition of the frequency response metric. Let $S_{wz,1}(t)$ and $S_{wz,2}(t)$ be the step responses of two systems from disturbance to performance output. For a set of sampling times $t_1, \dots, t_n$, we sample both step responses and vectorize to form vectors $s_1, s_2 \in \R^{n \cdot n_w \cdot n_z}$. Then, we define the step response metric as
\[
d_s(s_1, s_2) = \|s_1 - s_2\|_2.
\]

As with the frequency response metric, we find $L^p$ norms too computationally expensive.

\subsubsection{Perturbation Spectral Norm (PSN)}

While the frequency response and step response metrics defined above depend on the entire transfer function from disturbance to performance output, we note that the plants in our model only differ in their $A$ matrix. Therefore, for two plants $(A_1, B_1, C_1, D_1)$ and $(A_2, B_2, C_2, D_2)$ we define the perturbation spectral norm distance by
\[
d_p(A_1, A_2) = \|A_1 - A_2\|_2
\]
where \(\|\cdot\|_2\) is the spectral norm.


While this metric does not include the effects of the $B$, $C$, and $D$ matrices on the input-output transfer function, it is faster to compute than the other metrics mentioned.

\subsection{Clustering}

To group the $M$ contingencies into $k$ clusters using the selected metric, we consider several existing graph or metric clustering algorithms. These algorithms only require pairwise distances between points, not the points themselves (as might be required in, for example, k-means to average points), allowing the clustering algorithm to be chosen independently from the metric.

We consider three types of clustering: k-centers \cite{GONZALEZ1985293}, k-medoids \cite{pam}\cite{PARK20093336}, and divisive clustering \cite{Hastie2009}. These methods optimize for different objectives and have varying scalability.

\subsection{K-Centers}

For a set of $M$ points and a metric $d(i, j)$, the k-centers problem is to find clusters and centers of each cluster such that 
\[
    \max_i d(i, c(i))
\]
is minimized, where \(c(i)\) gives the index of the center of the cluster that point \(i\) is in. Note the center of a cluster is one of the points in the cluster. This method minimizes the maximum distance from a point to its cluster center.

While finding an optimal solution to this problem is NP-hard, there exist approximation algorithms with time complexity \(O(Mk)\) that can achieve a maximum distance to center within twice the optimal value \cite{GONZALEZ1985293}.

This clustering method is useful for robust group controller design techniques that view the radius of a group (the maximum distance between a point in the group and the center of the group) as the radius of an uncertainty set that covers plants in the group.

\subsection{K-Medoids}

The k-medoids problem is to find clusters and centers of each cluster to minimize
\[
\sum_i d(i, c(i)).
\]
Note that algorithms to minimize this objective can also be used to minimize \(\sum_i \tilde{d}(i, c(i))^2\) by setting \(d = \tilde{d}^2\). 

Algorithms of various approximation performance and time complexity are available, such as an algorithm with time complexity $O(k(M-k)^2)$ presented in \cite{pam} and one with time complexity $O(Mk)$ presented in \cite{PARK20093336}.

\subsection{Divisive Clustering}

Divisive clustering is an iterative clustering algorithm that starts with all points in one cluster and repeatedly divides the worst performing cluster until $k$ clusters are reached \cite{Hastie2009}. We choose the worst performing cluster as the one with the highest average distance of points in the cluster to the cluster center. This cluster is divided using another clustering algorithm, like one for k-centers or k-medoids.

While we divide clusters until we reach $k$ clusters, one might also divide until a certain performance criterion is achieved, if such a criterion is available and fast enough to compute.

\section{Simulation} \label{sec:simulation}


As case studies, we apply this grouping method to the IEEE 39-bus and 68-bus systems. We use the smaller 39-bus system to illustrate in detail the output of this grouping method. The case study of the 68-bus system demonstrates the scalability of the grouping method and also highlights that some of the observations on the IEEE 39-bus system are not universal.
{
We first demonstrate the grouping method with extended examples on both systems using $\mathcal{H}_\infty$ methods.
Then, we repeat the experiment on the IEEE 39-bus system using $\mathcal{H}_2$ control to demonstrate the applicability of our grouping method to different types of performance measurement.
}

\subsubsection{Distance Metrics}
When computing the distance between two contingencies, we put both perturbed systems in feedback with the nominal controller and apply the metric to these resulting closed-loop systems.
To sample the frequency responses for the frequency response metric, we first find the distance between the origin and all poles and zeros of both systems. Let $a$ be the smallest distance and $b$ be the largest distance. We then sample 1000 log-spaced frequencies in the interval $[\frac{a}{10}, 10b]$. For the step response metric, we take the absolute value of real parts of each pole of both system. Let the smallest value be $c$. We then sample 1000 time points in the interval $[0, \frac{1}{c}]$.

\subsubsection{Power System Modeling and Controller Design} \label{subsubsec:control-design}
{In all examples, the matrix $B_u$ is set to control of the generator exciters, where the excitation model used is Type AC4A from \cite{noauthor_ieee_2006}.  This represents control for voltage stability.} We also set $B_w = B_u$, so the disturbance enters into the control input. For solving power flows and computing linearized dynamics, we use the Power System Toolbox \cite{PST} and MatNetFlow libraries. All code was run in \textsc{Matlab} 2023a on a Intel Xeon Skylake 6230 at 2.1 GHz.

For control design, we use the \textsc{Matlab} \texttt{systune} function.
This function can be configured to design controllers that minimize the $\mathcal{H}_\infty$ norm or the $\mathcal{H}_2$ norm. 
{This function returns the parameters $A_k$, $B_{k x}$, $C_k$, and $D_{k x}$ of a controller in the form of equation~\ref{eq:controller}.
We configure \texttt{systune} to design a static $H_\infty$ (or $\mathcal{H}_2$) state feedback centralized controller with limited gains.
}
The gain limits reflect the lower and upper limits on the tunable parameters and can be adjusted independently for each parameter. We restrict the gains of each parameter to the interval $[-1, 1]$. This controller design method, applied to a group of plants $P_1, \dots, P_M$, designs the controller $K$ to minimize $\max_i \|\mathcal{F}(P_i, K)\|_{\mathcal{H}_\infty}$ (or $\max_i \|\mathcal{F}(P_i, K)\|_{\mathcal{H}_2}$) where \(\mathcal{F}(P_i, K)\) represents the feedback system of plant $P_i$ with controller $K$. Other methods that might be used for this control design process include the ones presented in \cite{xue2022} and \cite{MESANOVIC2020104490}.

\subsubsection{Performance Analysis} To analyze cost of a grouping $G = \{G_1, \dots, G_k\}$, we use comparisons with two extremes of grouping: the grouping $\{\{P_1, \dots, P_M\}\}$ where all contingencies are in the same group; and the grouping $\{\{P_1\}, \dots, \{P_M\}\}$ where each contingency in its own group (a group of size 1). Consider contingency $i$ corresponding to plant $P_i$.
Denote the group in grouping $G$ that contains contingency $i$ by $g(i; G)$ and the controller for a group $\tilde{G}$ by $K_{\tilde{G}}$. Then the controller applied to $P_i$ is $K_{g(i; G)}$.
Denote the controller for the group consisting of all contingencies by $K_w$, and the controller for the group consisting only of contingency $i$ by $K_i$.

Note that for any controller $K$ we have that $\|\mathcal{F}(P_i, K_i)\|_{\mathcal{H}_\infty} \leq \|\mathcal{F}(P_i, K)\|_{\mathcal{H}_\infty}$ since \(K_i\) by definition minimizes the \(\mathcal{H}_\infty\) norm when in feedback with $P_i$. Therefore, \(\|\mathcal{F}(P_i, K_i)\|_{\mathcal{H}_\infty} \leq \|\mathcal{F}(P_i, K_{g(i; G)})\|_{\mathcal{H}_\infty}\).
Further, since $g(i; G) \subseteq \{1, \dots, M\}$, we typically see that $\|\mathcal{F}(P_i, K_{g(i; G)})\|_{\mathcal{H}_\infty} \leq \|\mathcal{F}(P_i, K_w)\|_{\mathcal{H}_\infty}$. 




To compare cost across contingencies, we normalize a grouping's controller cost on each contingency, creating a scaled $\mathcal{H}_\infty$ norm greater than or equal to 0. The lower this scaled value, the higher the grouping's performance.
\begin{align}
    s(G, i) \triangleq \frac{\|\mathcal{F}(P_i, K_{g(i; G))})\|_{\mathcal{H}_\infty} - \|\mathcal{F}(P_i, K_i)\|_{\mathcal{H}_\infty}}{\|\mathcal{F}(P_i, K_w)\|_{\mathcal{H}_\infty} - \|\mathcal{F}(P_i, K_i)\|_{\mathcal{H}_\infty}}
    \label{eq:scaled-perf}
\end{align}

{
The numerator is the difference in cost between using the controller for contingency $i$ under grouping $G$, and using the optimal controller for contingency $i$.
The denominator is the difference in cost between using the controller designed to handle all contingencies on contingency $i$, and using the optimal controller for contingency $i$.
}

In this scale, a cost $s(G,i)$ of 0 represents a grouping $G$ performing optimally on contingency $i$, {meaning that the grouping $G$ results in the optimal controller for contingency $i$}.
A cost $s(G,i)$ of 1 represents a grouping $G$'s controller for contingency $i$ performing the same as the single controller designed to handle all contingencies.  Note that controllers may exist that score above 1 in this scale{, since it is possible to design worse performing controllers than $K_w$}.

As a measure of overall grouping cost, we average across all contingencies the cost of the grouping on each contingency. This is the mean scaled $\mathcal{H}_\infty$ norm, below:
\begin{align}
    \frac{1}{M} \sum_{i=1}^M s(G, i).
\end{align}

Note that the cost of a grouping depends both on the partition of contingencies and on the controller design method.

We note that an alternate, more time-consuming, method to analyze performance is through simulation of the full nonlinear dynamics. In practice, we may want to avoid even this in-depth linear performance analysis for the sake of computation time. One option is a quick performance verification procedure over each group using our recent dissipativity-based approach \cite{jensen2023certifying}.

\subsection{IEEE 39-bus System with $\mathcal{H}_\infty$ Control}
\label{subsec:ieee39-hinf}

\begin{figure}[t]
    \centering
    \includegraphics[width=\linewidth]{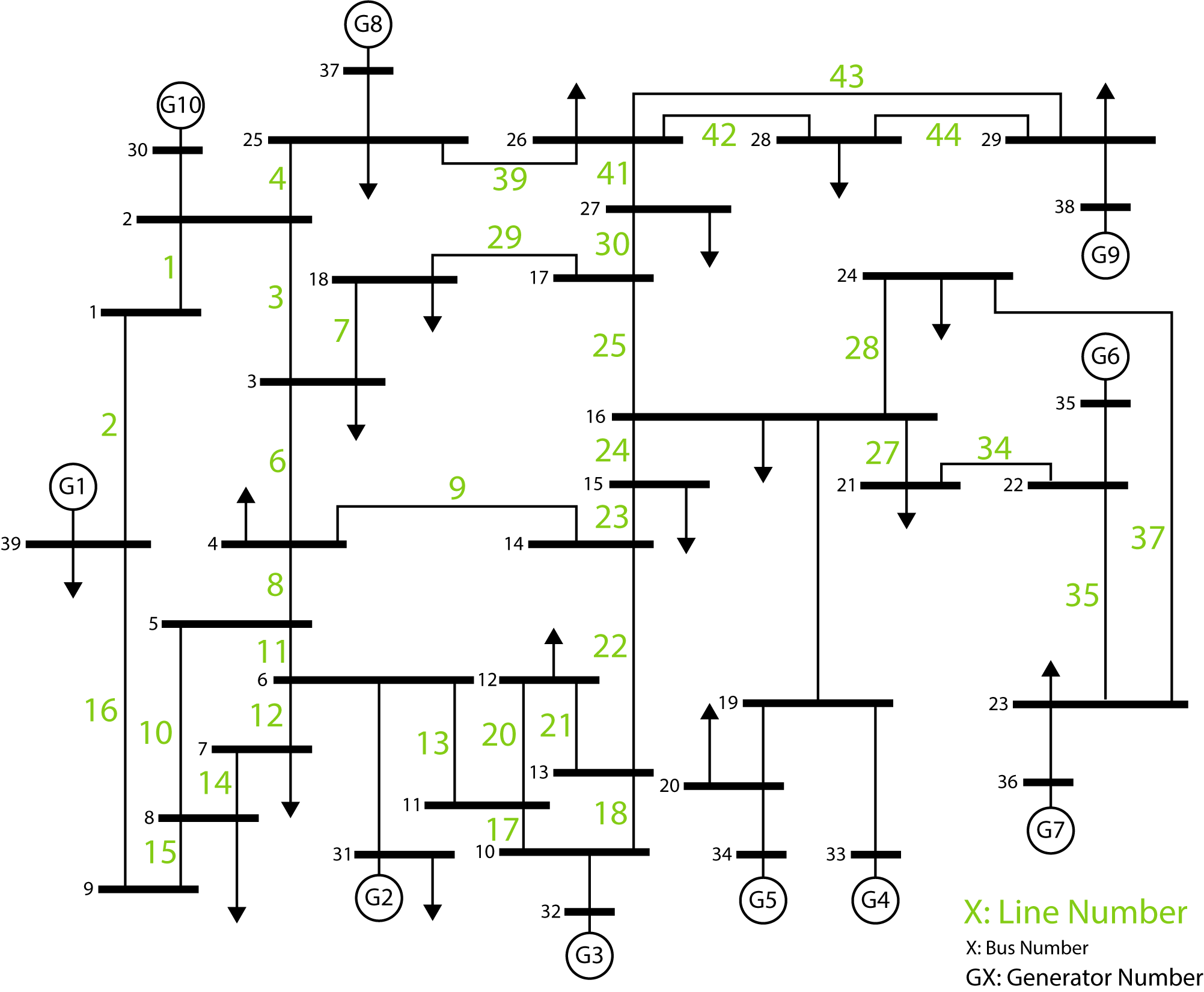}
    \caption{The IEEE 39-bus system with lines that may fail labeled. Only single-line failures that do not disconnect the grid are considered.}
    \label{fig:ieee39}
\end{figure}

The IEEE 39-bus system is a 10-machine, 39-bus, 46-line system representing the New England power system. Of the 46 lines, there are 11 whose removal will disconnect the underlying graph. We assume these line failures are handled separately, and only consider the remaining $M=35$ lines as contingencies. These lines are labeled in Fig. \ref{fig:ieee39}. In this example we study the performance of the algorithm presented in Section \ref{sec:grouping} with several metrics and clustering methods, for varying numbers of groups.

\begin{figure*}[t]
    \centering
    \includegraphics[width=0.7\linewidth,trim={0.1in 0.0in 0.6in 0in},clip]{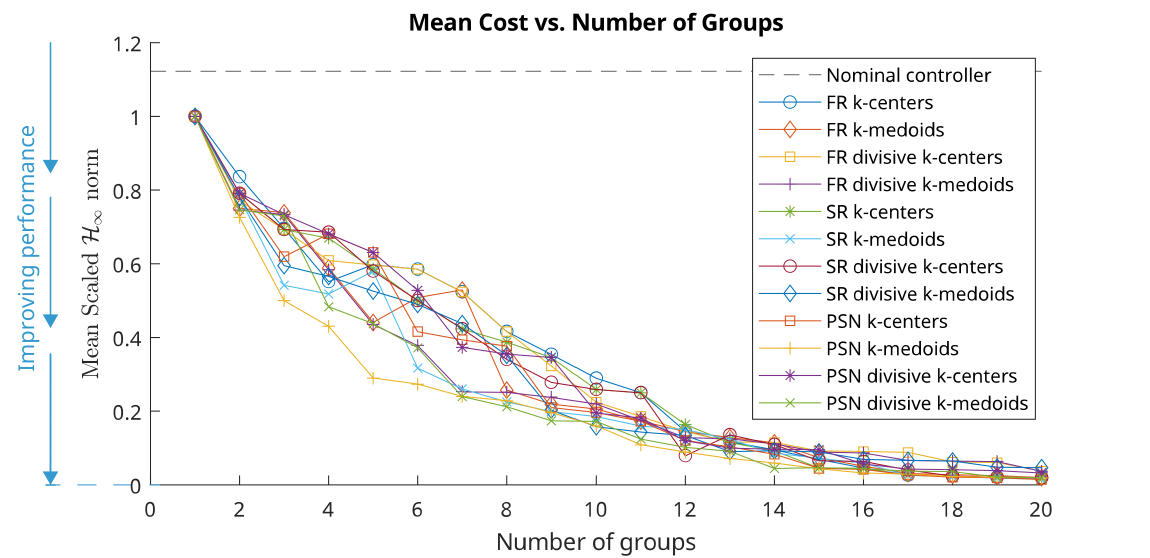}
    \caption{Comparison of metrics and clustering algorithms on IEEE 39-bus system by scaled $\mathcal{H}_\infty$ norm averaged over contingencies versus number of groups. FR is the frequency response metric, SR is the step response metric, and PSN is the perturbation spectral norm metric.}
    \label{fig:ieee39-perf-vs-num-groups}
\end{figure*}

We run Algorithm~\ref{alg:alg} on all nine combinations of the metric and clustering methods in Section~\ref{sec:grouping} for numbers of groups $k = 1, \dots, 20$ and compare performance across metric and clustering methods versus number of groups. Figure~\ref{fig:ieee39-perf-vs-num-groups} summarizes the average scaled $\mathcal{H}_\infty$ norm over contingencies versus number of groups for each grouping method. As a baseline, we include the cost of the nominal controller applied to the contingencies. All tested grouping methods perform near optimal on every contingency when partitioning the contingencies into 20 groups. For numbers of groups under 10, we observe four well-performing methods: PSN k-medoids, PSN divisive k-medoids, SR k-medoids, and FR divisive k-medoids. Additionally, for this system, we observe that designing just one controller to handle all contingencies performs better on average than continuing to use the nominal controller after a contingency occurs. 

\begin{figure}
    \centering
    \includegraphics[width=\linewidth, trim={0.3in 0.1in 0.2in 0.1in},clip]{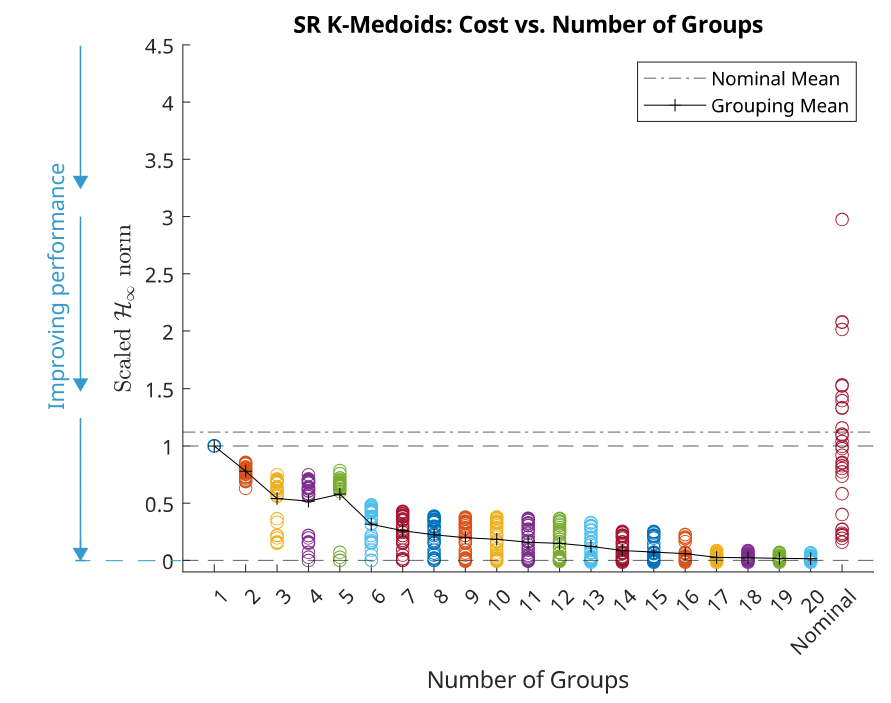}
    \caption{Scaled $\mathcal{H}_\infty$ norm of each contingency when applying the step response metric and k-medoids clustering algorithm for various numbers of groups. The ``nominal'' column is the scaled $\mathcal{H}_\infty$ norm when applying the nominal controller to each contingency.}
    \label{fig:ieee39-sr-kmd-perf-vs-num-groups}
\end{figure}

While Figure~\ref{fig:ieee39-perf-vs-num-groups} summarizes mean cost of grouping algorithms, mean cost may not be the only performance characteristic of interest.
Figure~\ref{fig:ieee39-sr-kmd-perf-vs-num-groups} plots the scaled closed-loop $\mathcal{H}_\infty$ norm for each contingency when applying the step response metric with k-medoids clustering for numbers of groups 1 through 20. This shows the distribution of the cost on each contingency for this grouping method for groupings into various numbers of groups. The right-most column displays the scaled closed-loop $\mathcal{H}_\infty$ norm when applying the nominal controller to each contingency. There are several contingencies on which the nominal controller achieves a scaled $\mathcal{H}_\infty$ norm of under 0.5. There are also several contingencies on which the nominal controller achieves a scaled $\mathcal{H}_\infty$ norm of over 1.0, which is worse than the performance achieved by one controller designed to handle all contingencies. For low numbers of groups, the scaled $\mathcal{H}_\infty$ norms achieved by the step response and k-medoids grouping method does not decrease consistently across contingencies. Instead, some contingencies see a large decrease in scaled $\mathcal{H}_\infty$ norm while the performance on other contingencies improves only slightly.

\begin{figure}[t]
    \centering
    \includegraphics[width=\linewidth, trim={0.3in 0.0in 0.1in 0.4in},clip]{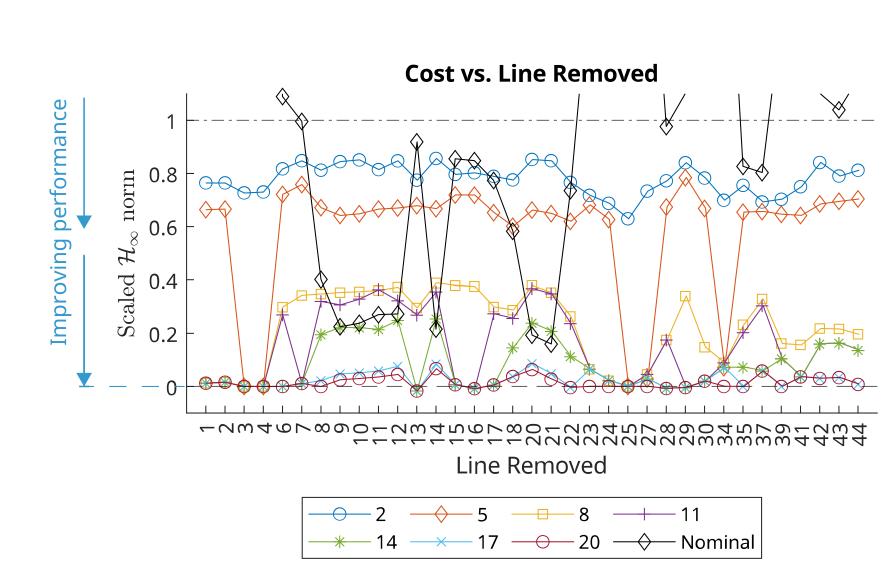}
    \caption{Scaled $\mathcal{H}_\infty$ norm of each contingency for various numbers of groups using step response metric and k-medoids clustering. The scaled $\mathcal{H}_\infty$ norm of the nominal controller is cutoff above 1.1.}
    \label{fig:ieee39-sr-kmd-perf-vs-line-removed}
\end{figure}

Figure~\ref{fig:ieee39-sr-kmd-perf-vs-line-removed} shows the change in cost on each contingency as the number of groups increases when using the step response metric and k-medoids clustering algorithm. The cost of the nominal controller is also shown, but cutoff for those contingencies where the nominal controller's scaled $\mathcal{H}_\infty$ norm is above 1.1 since, as seen in Figure~\ref{fig:ieee39-sr-kmd-perf-vs-num-groups}, the nominal controller's scaled $\mathcal{H}_\infty$ norm goes as high as 4.5 for one contingency on this system. As expected, the cost on any particular contingency is generally non-increasing in the number of groups. Across most contingencies, there is a large improvement in performance between 5 and 8 groups, and improvements in performance become more incremental as the number of groups rises after that. Since additional groups means additional computation time to design the extra controllers, this information could be used by an operator to refine the number of groups to use in the future. The scaled $\mathcal{H}_\infty$ norms of some lines such as 3, 4, 25, 27, and 34 drops to near-optimal after 2 groups, suggesting these lines are split into smaller groups by the clustering algorithm at low numbers of groups. Note that while the scaled $\mathcal{H}_\infty$ norm of the nominal controller on some contingencies is high enough that it is not displayed in this figure, on a few contingencies such as lines 8, 9, 10, 11, 13, 18, and 19, the nominal controller achieves a level of performance that exceeds that of the step response and k-medoids grouping method at 11 groups. This suggests one direction to improve this grouping algorithm would be to use the nominal controller itself as the basis for one group, since the nominal controller is pre-computed.

\begin{figure}[t]
    \centering
    \includegraphics[width=\linewidth, trim={0.2in, 2.5in, 0.6in, 2.2in},clip]{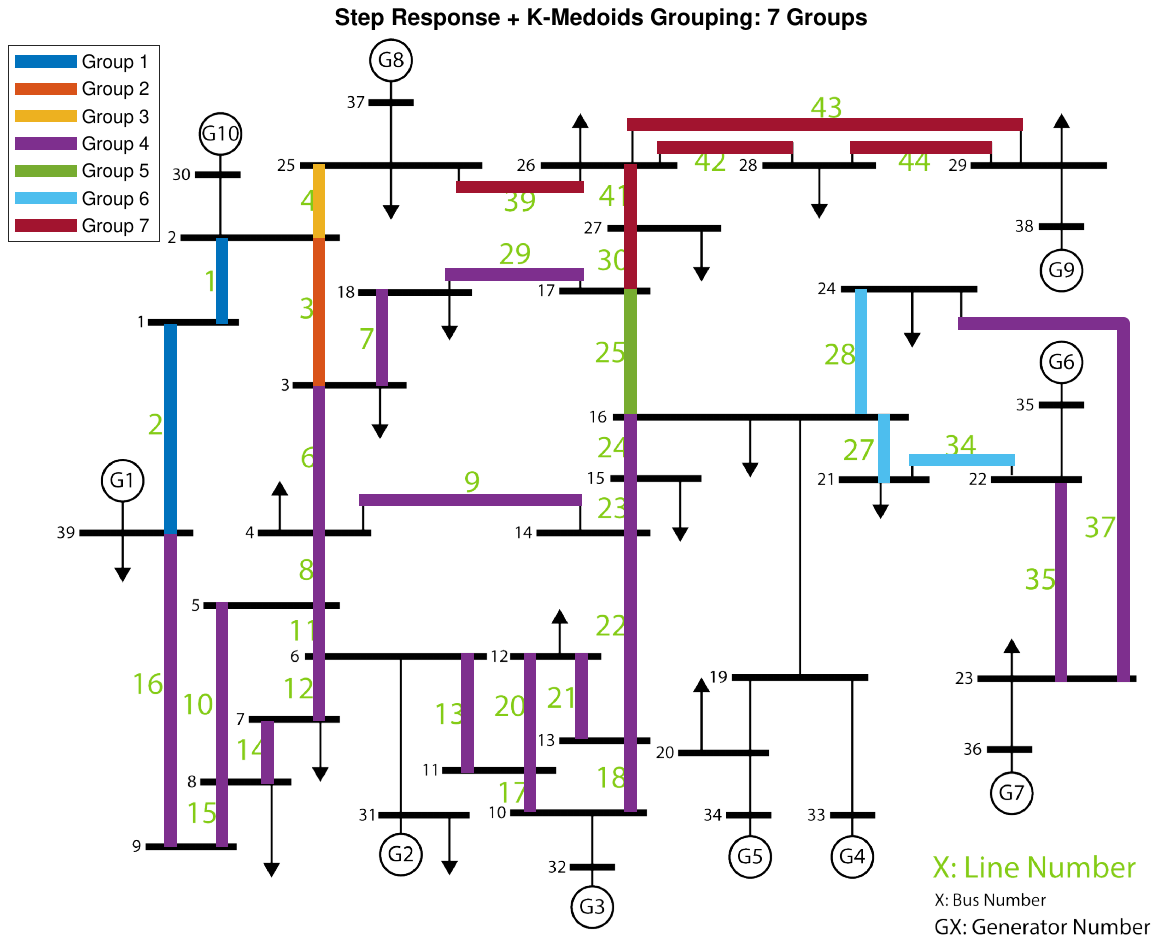}
    \caption{The grouping generated using the step response metric and k-medoids clustering for 7 groups. Lines colored the same are in the same group. One controller is designed per group. If any line in the group fails, then that controller is applied. }
    \label{fig:ieee39-sr-kmd-grouping-7groups}
\end{figure}

\begin{figure}[t]
    \centering
    \includegraphics[width=\linewidth, trim={0.2in, 2.5in, 0.6in, 2.2in},clip]{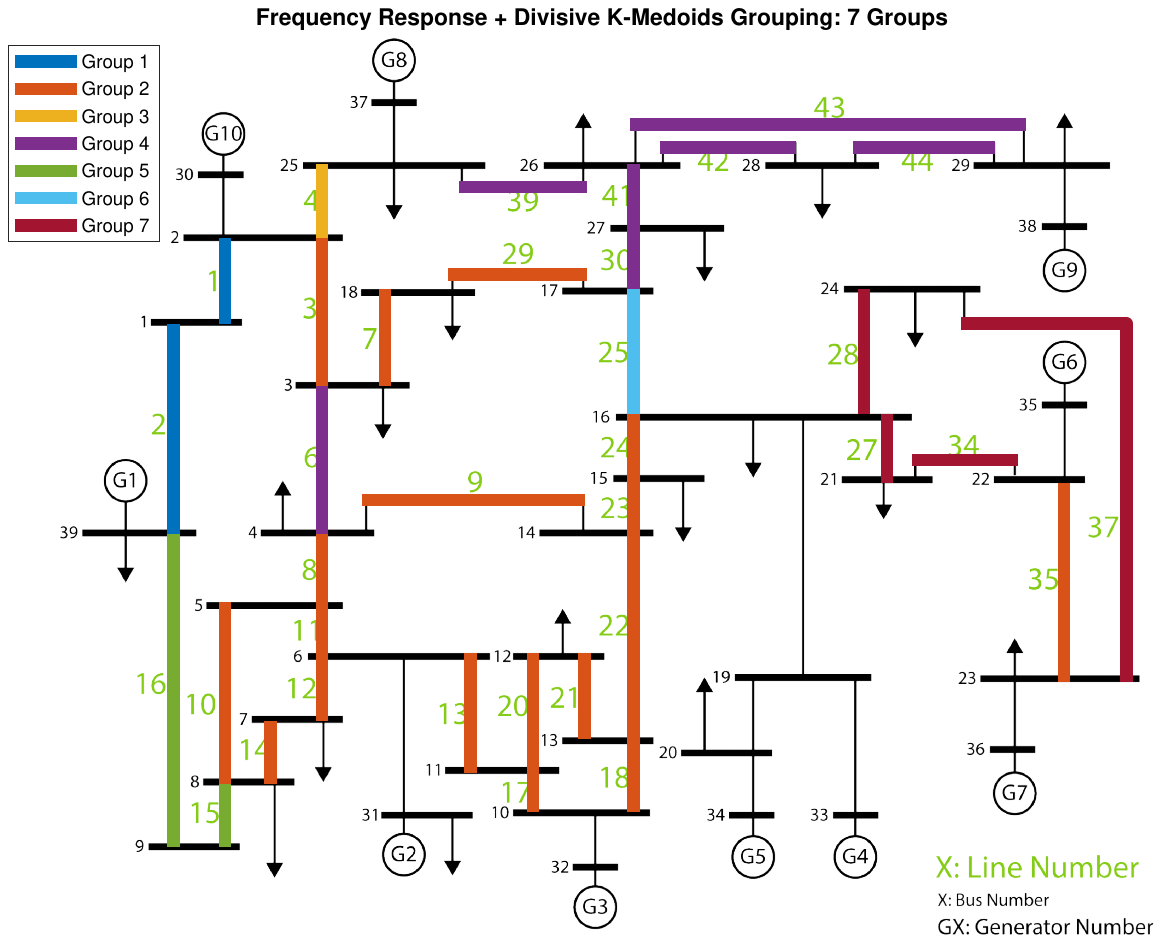}
    \caption{The grouping generated using the frequency response metric and divisive k-medoids clustering for 7 groups. Lines colored the same are in the same group.}
    \label{fig:ieee39-fr-dkmd-grouping-7groups}
\end{figure}

For illustration of the groupings generated by this grouping method, we examine two groupings in closer detail: the grouping generated using the step response metric, k-medoids clustering, and a choice of 7 groups (depicted in Figure~\ref{fig:ieee39-sr-kmd-grouping-7groups}); and the grouping generated using the frequency response metric, divisive k-medoids clustering, and a choice of 7 groups (depicted in Figure~\ref{fig:ieee39-fr-dkmd-grouping-7groups}). These two groupings have similar mean scaled $\mathcal{H}_\infty$ norm cost.

While the two groupings are generated using both different metrics and different clustering algorithms, there are several similarities between the groupings. Both have a group consisting of only lines 1 and 2. Group 7 of the step response \& k-medoids grouping consists of lines 30, 39, 41, 42, 43, and 44, and group 4 of the frequency response \& divisive k-medoids grouping is almost the same, only differing through the addition of line 6. Both groupings also have a  ``catch-all'' group: group 4 in the step response \& k-medoids grouping, and group 2 in the frequency response \& divisive k-medoids grouping. Finally, both groupings isolate lines 4 and 25 into groups of size 1.

Compared to the nominal controller, both of these groupings improve significantly upon the highest $\mathcal{H}_\infty$ (not scaled) norms: the highest $\mathcal{H}_\infty$ norm seen using the nominal controller is $0.202$, when applied to line 25, while under the step response \& k-medoids grouping it is $0.152$, when applied to line 37, and under the frequency response \& divisive k-medoids grouping it is $0.152$, when applied to line 3. The five highest $\mathcal{H}_\infty$ norms seen using the nominal controller are 0.202, 0.190, 0.181, 0.179, and 0.177 on lines 25, 4, 3, 34, and 27 respectively. These same contingencies see the $\mathcal{H}_\infty$ norms of 0.151, 0.149, 0.146, 0.149, and 0.149 respectively using the step response \& k-medoids grouping, and $\mathcal{H}_\infty$ norms of 0.151, 0.149, 0.152, 0.149, and 0.149 respectively using the frequency response \& divisive k-medoids grouping. Under both groupings, the $\mathcal{H}_\infty$ norms on these lines is reduced significantly.

Some aspects of the groupings are intuitive. For example, lines 1 and 2 are in the same group in both groupings, which is unsurprising since the two lines are connected by a bus with no other lines, loads, or generators. Additionally, many groups in both groupings consist of lines that are close together in the power system. In fact, only group 4 in the step response \& k-medoids grouping, and groups 2 and 4 in the frequency response \& divisive k-medoids are disconnected. It is these disconnected groups where the utility of this automated grouping method is demonstrated. It may not be obvious to a human operator that line 35 might be put with lines 7, 8, 9, etc., as done in group 4 of the step response \& k-medoids grouping, instead of with lines 27, 28, and 34. We also note that several lines are isolated into groups of size one by each grouping: the step response \& k-medoids grouping isolates lines 3, 4, and 25; and the frequency response \& divisive k-medoids grouping isolates lines 4 and 25. While line 3 is not isolated in the frequency response \& divisive k-medoids grouping, it is the line on which this grouping performs the worst. This suggests that the removal of any of lines 3, 4, or 25 causes unique changes to the power system dynamics. This type of analysis could be useful to a power system operator for identifying contingencies which may need further consideration.

Table~\ref{table:ieee39-metric-times} summarizes the average time required to evaluate each metric once an Intel Xeon Skylake 6230 at 2.1 GHz. Running the clustering algorithm may require all \(M (M-1) / 2\) pairwise distances. For the IEEE 39-bus system, this means computing all pairs of distances on a single thread using the frequency response metric takes 55.63 seconds, using the step response metric takes 28.95 seconds, and using the perturbation spectral norm metric takes 0.1039 seconds. This can be parallelized to bring the computation time for the distance matrix down by a constant factor. 
\begin{table}[b]
    \centering
    \caption{Average time to compute distance between two contingencies on the IEEE 39-bus system for various metrics.}
    \begin{tabular}{|c|c|c|}
        \hline 
        \multicolumn{3}{|c|}{Average Metric Computation Time (s)} \\
        \hline
        FR & SR & PSN \\
        \hline
        0.0935 & 0.0487 & 0.000175 \\
        \hline
    \end{tabular}
    \label{table:ieee39-metric-times}
\end{table}

{
Over all controllers synthesized for these IEEE 39-bus system experiments, the time to synthesize a single controller ranged between a minimum of 33.771 seconds and a maximum of 512.746 seconds, with a mean of 138.864 seconds.
With these times, Algorithm~\ref{alg:alg} could be re-run---using appropriate parallelism for control design---every 15 minutes to account for changing load and generation levels, even with the worst case controller design times.
Note that this is heavily dependent on the controller design algorithm used, and in these experiments we used methods readily available in \textsc{Matlab}.
Other works such as \cite{MESANOVIC2020104490} and \cite{xue2022} present methods for computationally efficient $\mathcal{H}_\infty$ controller design.
}

\subsection{IEEE 68-bus System with $\mathcal{H}_\infty$ Control}
\label{subsec:ieee68-hinf}

We also apply our grouping method to the larger IEEE 68-bus system.
Of note, the perturbation spectral norm metric does not perform well on this system, even though it did on the 39-bus system.
This 68-bus system, extracted from data presented in \cite{osti_6712007}, consists of 16-machines, 68 buses, and 86 lines. For further description of this system, we refer to \cite{wu_decentralized_2015}. Of the 86 lines in this system, there are 18 lines whose removal disconnects the graph, and 4 more lines whose removal results in a lack of a power flow solution. We exclude these lines and consider only the remaining $M=64$ lines as contingencies.

\begin{figure}[t]
    \centering
    \includegraphics[width=\linewidth,trim={0.1in 0in 0in 0in},clip]{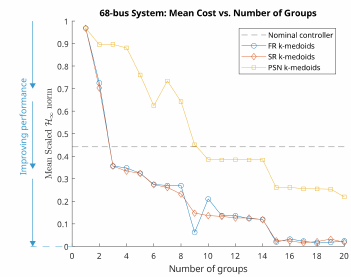}
    \caption{Comparison of metrics using the k-medoids clustering algorithm on the IEEE 68-bus system by scaled $\mathcal{H}_\infty$ norm averaged over contingencies versus number of groups. FR is the frequency response metric, SR is the step response metric, and PSN is the perturbation spectral norm metric.}
    \label{fig:68-perf-vs-num-groups}
\end{figure}

We compute groupings using the step response, frequency response, and perturbation spectral norm metrics with the k-medoids clustering algorithm based on the results on the IEEE 39-bus system and compare the mean cost versus the number of groups in Figure~\ref{fig:68-perf-vs-num-groups}. The mean scaled $\mathcal{H}_\infty$ norm approaches 0 (optimal) for all three methods faster relative to the number of contingencies than they did on the IEEE 39-bus system. However, while the perturbation spectral norm metric performed better than the other metrics on the IEEE 39-bus system, it significantly under-performs compared to the step and frequency response metrics on this system. Both the step response and frequency response metrics with k-medoids outperform the nominal controller on average over the contingencies with just 3 groups and perform close to optimally on average at 15 or more groups. This is a significant reduction from the 64 controllers that would need to be designed to handle each contingency individually.

\begin{figure}
    \centering
    \includegraphics[width=\linewidth,trim={0.3in 0in 0.1in 0.1in},clip]{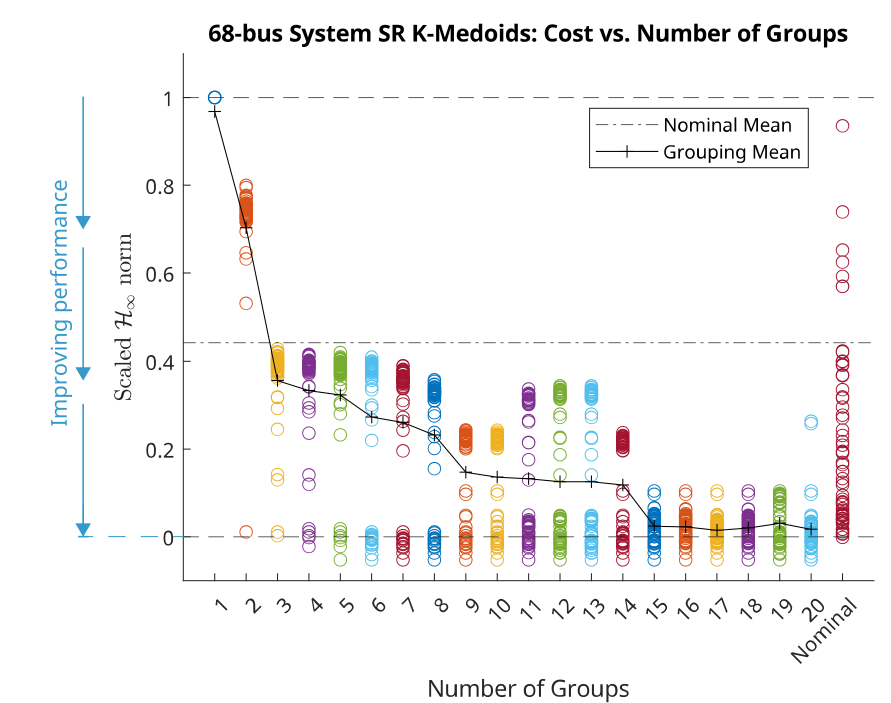}
    \caption{Scaled $\mathcal{H}_\infty$ norm when applying the step response metric and k-medoids clustering algorithm on the IEEE 68-bus system for various numbers of groups. The ``nominal'' column is the scaled $\mathcal{H}_\infty$ norm when applying the nominal controller to each contingency. Note that the $y$-axis is limited to $[-0.1, 1.1]$; the nominal controller's values go up to 7.02. }
    \label{fig:68-sr-kmd-perf-vs-num-groups}
\end{figure}

Figure~\ref{fig:68-sr-kmd-perf-vs-num-groups} shows the cost applying the step response and k-medoids grouping method to each contingency in the IEEE 68-bus system for various numbers of groups. As in the IEEE 39-bus system, the grouping method tends to create groups on which controllers do near-optimal, and a remaining large group whose performance slowly improves as the number of groups increases. Note that while the scaled $\mathcal{H}_\infty$ norm should be greater than or equal to 0, we see some slightly negative values due to controller design variance in \texttt{systune}.

The five highest $\mathcal{H}_\infty$ norms (not scaled) using the nominal controller on each contingency in this system are 0.262, 0.249, 0.148, 0.138, and 0.130. Under the grouping into 10 groups with the step response and k-medoids grouping algorithm, these $\mathcal{H}_\infty$ norms become 0.120, 0.123, 0.101, 0.102, and 0.109 respectively. The overall highest $\mathcal{H}_\infty$ norm seen under this grouping is 0.123. This is less than half of the worst-case $\mathcal{H}_\infty$ norm from continued use of the nominal controller.

Due to the larger state dimension when modeling this system, the time required to compute the distance between two contingencies increases significantly: the frequency response metric takes 0.653 seconds to evaluate compared to 0.0935 seconds on the IEEE 39-bus system; the step response metric takes 0.3522 seconds compared to 0.0487 seconds on the IEEE 39-bus system; and the perturbation spectral norm metric takes 0.000959 seconds compared to 0.000175 seconds on the IEEE 39-bus system. While the perturbation spectral norm is fast to compute, its performance does not match that of the other metrics tested on the IEEE 68-bus system. 

{
\subsection{IEEE 39-bus System with $\mathcal{H}_2$ Control}
}

{
In this section, we repeat the experiments of Section~\ref{subsec:ieee39-hinf} using
$\mathcal{H}_2$ control instead of $\mathcal{H}_\infty$ control.
To design the controllers, we use \texttt{systune} configured for $\mathcal{H}_2$ control. The rest of the control design configuration is the same as in Section~\ref{subsubsec:control-design}.
Other than the change in control design and performance measurement, our testing methodology is the same as in Section~\ref{subsec:ieee39-hinf}.
Note that these changes do not affect the distance metrics or grouping.
}

\begin{figure}
    \centering
    \includegraphics[width=\linewidth]{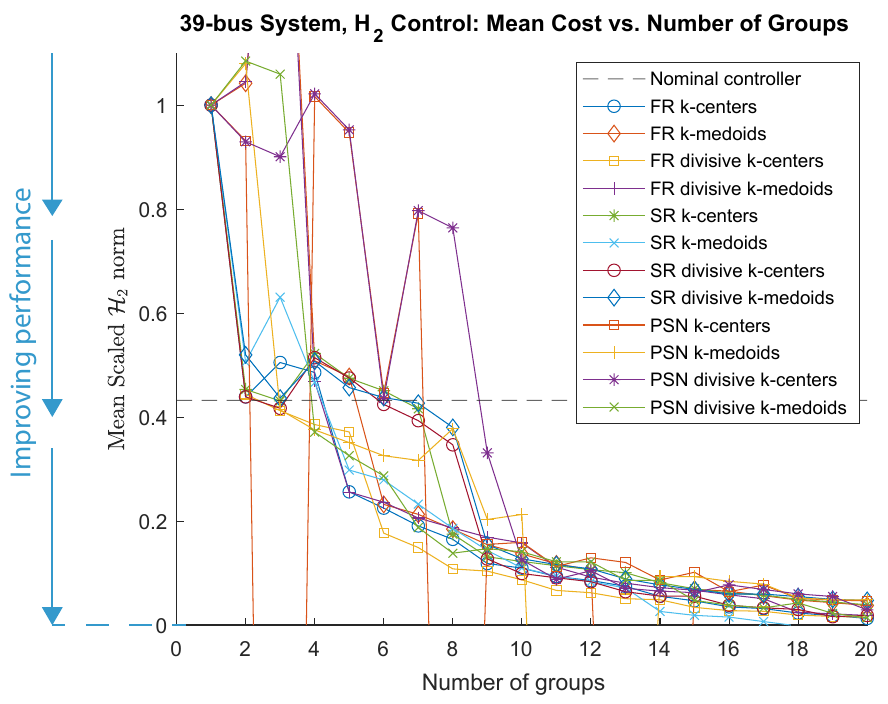}
    \caption{Comparison of metrics and clustering algorithms on IEEE 39-bus system by scaled $\mathcal{H}_2$ norm averaged over contingencies versus number of groups. FR is the frequency response metric, SR is the step response metric, and PSN is the perturbation spectral norm metric.}
    \label{fig:h2-ieee39-mean-cost-vs-num-groups}
\end{figure}

{
Figure~\ref{fig:h2-ieee39-mean-cost-vs-num-groups} summarizes the average scaled $\mathcal{H}_2$ norm over contingencies versus number of groups for each grouping method.
The SR k-medoids, FR k-medoids, FR divisive k-medoids, FR k-centers, FR divisive k-centers, and PSN divisive k-medoids methods perform the best for numbers of groups under 9. These methods outperform the nominal controller at 4 or more groups. All methods perform roughly the same at 10 or more groups. The PSN metric, with both k-centers and divisive k-centers, performed erratically, not achieving a monotonic decrease in mean scaled $\mathcal{H}_2$ norm that would be desired.
}

\begin{figure}
    \centering
    \includegraphics[width=\linewidth]{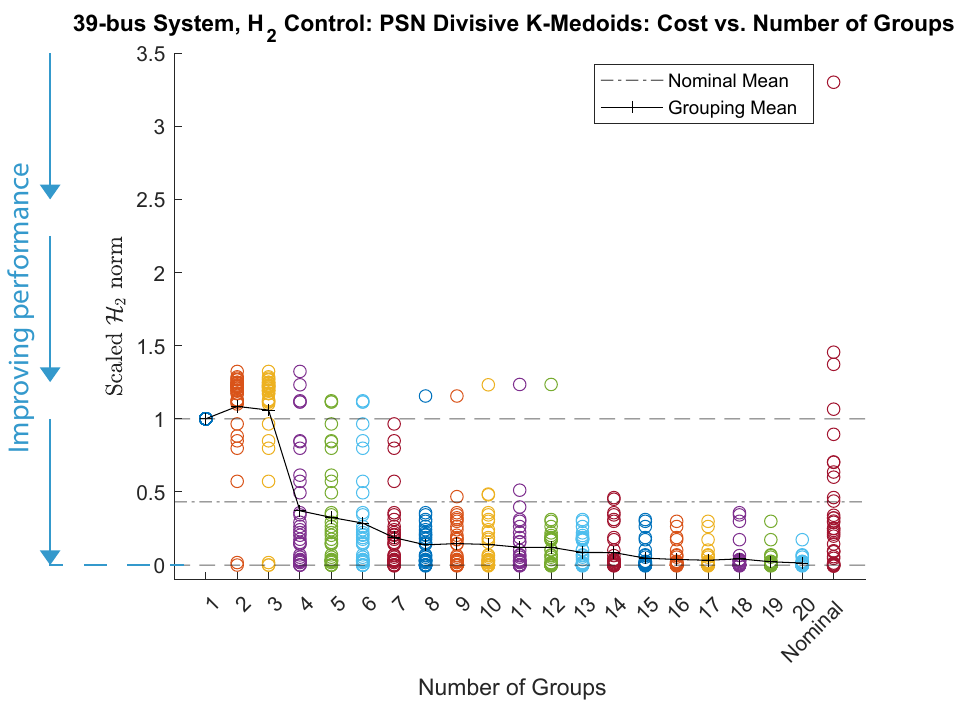}
    \caption{Scaled $\mathcal{H}_2$ norm of each contingency when applying the perturbation spectral norm metric and divisive k-medoids clustering algorithm for various numbers of groups. The ``nominal'' column is the scaled $\mathcal{H}_2$ norm when applying the nominal controller to each contingency.}
    \label{fig:h2-ieee39-psndkmd-vs-num-groups}
\end{figure}

{
Figure~\ref{fig:h2-ieee39-psndkmd-vs-num-groups} plots the scaled closed-loop $\mathcal{H}_2$ norm for each contingency when applying the perturbation spectral norm metric with divisive k-medoids clustering for various numbers of groups.
For two and three numbers of groups, the scaled $\mathcal{H}_2$ norms are above 1, indicating variance in the results of the control design method.
The mean drops rapidly between from three to four groups, but the performance on many individual contingencies remains high until eight groups and above.
Between eight and twelve groups, the groupings perform poorly on one contingency.
Further analysis shows that this contingency corresponds to the removal of line 3 in all of these groupings. The removal of line 3 was found to cause unique changes to the power system dynamics in Section~\ref{subsec:ieee39-hinf}.
}

\vspace{1em}
\section{Conclusion \& Future Work} \label{sec:conclusion}

In this paper we presented a method to improve the robustness of a power system in the event of a contingency. Our method involves automating the process of determining groups of contingencies that result in similar dynamics. This is used to partition the set of contingencies into a smaller set of groups of contingencies. We showed this finds high performing groupings that may not be identified by human operators, and can also be used to identify severe contingencies.
We further showed that by designing one controller for each group of contingencies, we can reduce the number of controllers that need to be designed to handle all contingencies, resulting in computational savings.
This method naturally allows trading-off between computation time and performance by choosing the number of groups into which the set of contingencies is partitioned{, enabling operators to choose a balance for their specific system and computational resources}.
We demonstrated the effectiveness of this approach in simulation.

One direction of future work is in developing improved metrics to determine similarity between contingencies. A metric with low computation time is important since the number of distance computations scales in the square of the number of contingencies. A good metric should also scale well in the dimension of the system. One approach may be to extract features important for similarity from the frequency response and compare just these features.

While in this paper we presented control design as largely independent of grouping, another direction is to tie control design to the particular metric and clustering algorithm being used. The metric might be used as a measure of uncertainty and the clustering algorithm may aim to create groups that represent small uncertainty sets. A robust control design procedure for these uncertainty sets could allow greater guarantees of overall system robustness. The k-centers algorithm or a sequential-greedy version of it may be useful for this, since it aims to reduce the radii of groups.

Other directions of future work are: leveraging performance criteria to automatically determine the number of groups into which the set of contingencies should be partitioned; generalizing this grouping method to handle other types of contingencies such as generator failures, transformer failures{, and $N-k$ contingencies; and exploring the effect of distributed energy resources on this type of clustering method}.

\printbibliography

\begin{IEEEbiography}[{\includegraphics[width=1in,height=1.25in,clip,keepaspectratio]{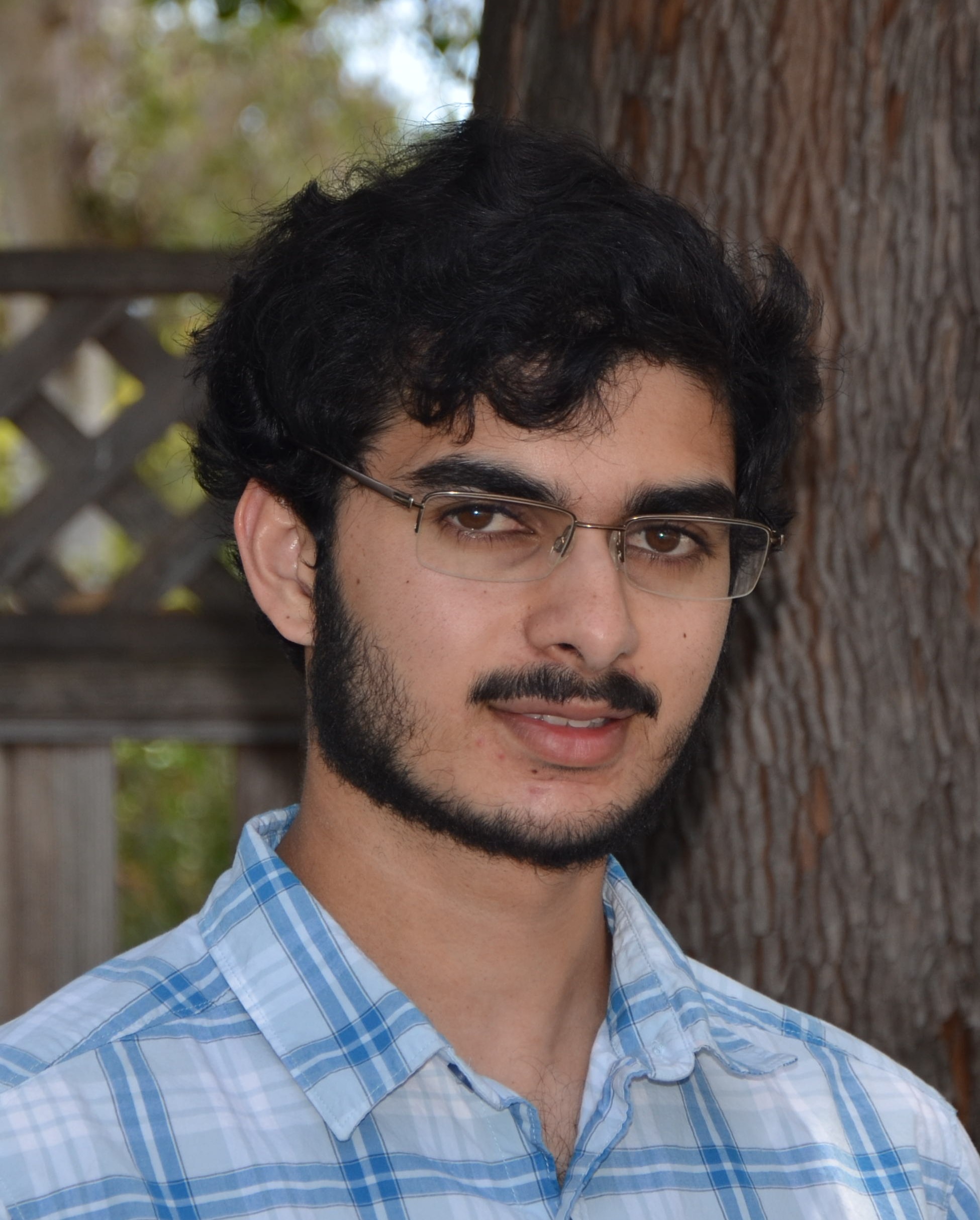}}]{Neelay Junnarkar} is a Ph.D. student at UC Berkeley and previously received a B.S. in Electrical Engineering and Computer Sciences from UC Berkeley (2021). He worked as an intern with Siemens Technology in 2023. His research interests include applications of machine learning to control theory.
\end{IEEEbiography}

\begin{IEEEbiography}[{\includegraphics[width=1in,height=1.25in,clip,keepaspectratio]{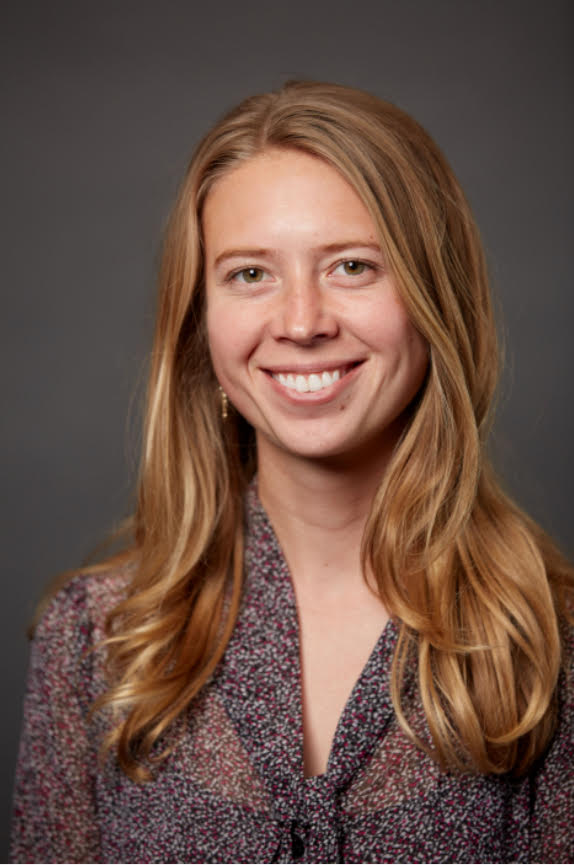}}]{Emily Jensen} is a postdoctoral researcher at UC Berkeley and previously held a postdoctoral appointment at Northeastern University. She
received her B.S. in Engineering, Mathematics \& Statistics from UC Berkeley in 2015 and the Ph.D. degree in Electrical
\& Computer Engineering from the University of California,
Santa Barbara in 2020. She was a research \& instructional assistant at Caltech in 2016. Dr. Jensen received the UC Regents’ Graduate Fellowship (2016), the Zonta Amelia Earhart Fellowship (2019) and an honorable mention for the Young Author
Award at the IFAC Conference on Networked Systems (2022). 
\end{IEEEbiography}
 
\begin{IEEEbiography}[{\includegraphics[width=1in,height=1.25in,clip,keepaspectratio]{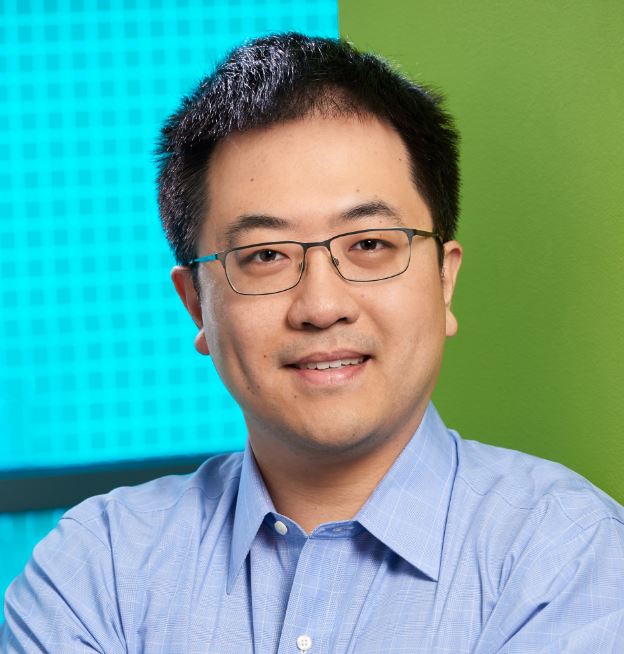}}]{Xiaofan Wu} received the B.Eng. degree in detection guidance and control technology from the Beijing University of Aeronautics and Astronautics, Beijing, China, in 2010, and the M.S. and Ph.D. degrees in electrical engineering from the University of Minnesota, Minneapolis, MN, USA, in 2012 and 2016, respectively. He is currently the Head of Research Group - Autonomous System and Control with Siemens Technology in Princeton, NJ, USA. He is also a Project Manager for the Siemens Princeton Island Grid Living Lab. He has more than ten years of experience in modeling, control and optimization of energy systems and smart grids. He was a Visiting Scholar with the Automatic Control Laboratory, ETH Zurich, Switzerland, from March to June 2016. He was the recipient of the 2021 Thomas Edison Patent Award of the New Jersey R\&D Council.
\end{IEEEbiography}

\begin{IEEEbiography}[{\includegraphics[width=1in,height=1.25in,clip,keepaspectratio]{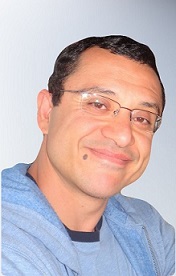}}]{Suat Gumussoy} is a senior expert on data-driven control at Autonomous Systems \& Control group at Siemens Technology in Princeton, NJ. His research interests are learning, control, and autonomous systems with particular focus on reinforcement learning, data-driven modeling and control, and development of commercial engineering design tools.

Dr. Gumussoy received his B.S. degrees in Electrical \& Electronics Engineering and Mathematics from Middle East Technical University at Turkey (1999) and M.S., Ph.D. degrees in Electrical and Computer Engineering from The Ohio State University at USA (2001, 2004). He worked as a system engineer in defense industry (2005-2007), as a postdoctoral associate in Computer Science Department at KU Leuven (2008-2011), and a principal control system engineer in Controls \& Identification Team at MathWorks where his contributions ranges from numerical algorithms to comprehensive analysis \& design tools in Control System, Robust Control, System Identification and Reinforcement Learning Toolboxes.

He served as an Associate Editor in IEEE Transactions on Control Systems Technology and IEEE Conference Editorial Board for 2018-2022. He is a Senior Member of IEEE and a member of IFAC Technical Committee on Linear Control Systems.
\end{IEEEbiography}

\begin{IEEEbiography}[{\includegraphics[width=1in,height=1.25in,clip,keepaspectratio]{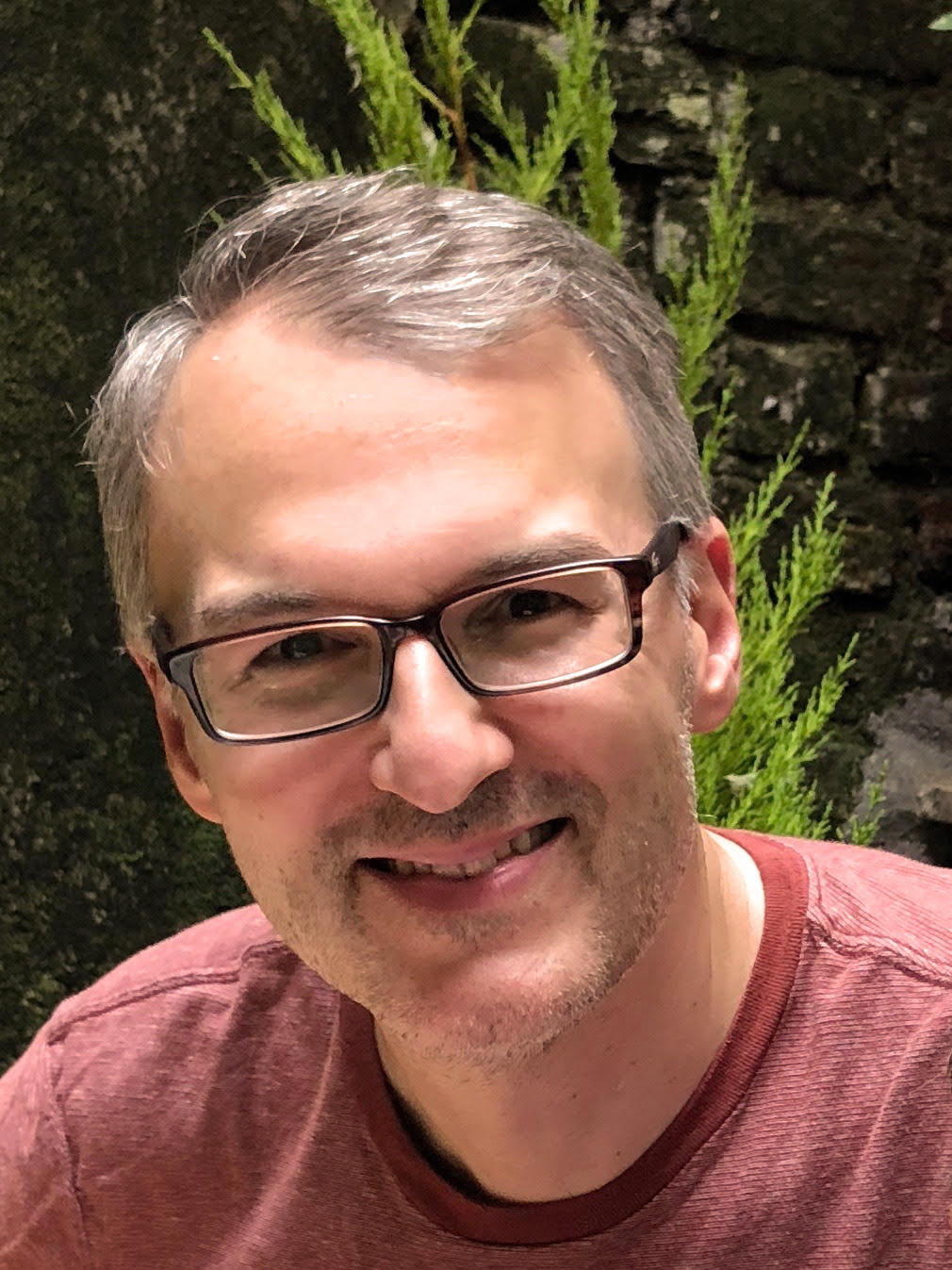}}]{Murat Arcak} is a professor at U.C. Berkeley in the Electrical Engineering and Computer Sciences Department, with a courtesy appointment in Mechanical Engineering.  He received the B.S. degree in Electrical Engineering from the Bogazici University, Istanbul, Turkey (1996) and the M.S. and Ph.D. degrees from the University of California, Santa Barbara (1997 and 2000). He received a CAREER Award from the National Science Foundation in 2003, the Donald P. Eckman Award from the American Automatic Control Council in 2006, the Control and Systems Theory Prize from the Society for Industrial and Applied Mathematics (SIAM) in 2007, and the Antonio Ruberti Young Researcher Prize from the IEEE Control Systems Society in 2014. He is a member of ACM and SIAM, and a fellow of IEEE and the International Federation of Automatic Control (IFAC).
\end{IEEEbiography}

\end{document}